# Comparison of Measured and Calculated Specific Resistances of Pd/Pt Interfaces.

S.K. Olson, R. Loloee, N. Theodoropoulou, W.P. Pratt Jr. and J. Bass
*Department of Physics, Center for Fundamental Materials Research, and Center for Sensor Materials, Michigan State University, East Lansing, MI 48824-1116*

P.X. Xu and Ke Xia
*State Key Laboratory for Surface Physics, Institute of Physics, Chinese Academy of Sciences, Beijing China*

We compare specific resistances (AR = area A times resistance R) of sputtered Pd/Pt interfaces measured in two different ways with no-free-parameter calculations. One way gives $2AR_{Pd/Pt}$ = 0.29 ± 0.03 f$\Omega$m$^2$ and the other $2AR_{Pd/Pt}$ = 0.17 ± 0.13 f$\Omega$m$^2$. From these we derive a 'best estimate' of $2AR_{Pd/Pt}$ = 0.28 ± 0.06 f$\Omega$m$^2$, which overlaps with no-free-parameters calculations: $2AR_{Pd/Pt}$ (predicted) = 0.30 ± 0.04 f$\Omega$m$^2$ for flat, perfect interfaces, or 0.33 ± 0.04 f$\Omega$m$^2$ for interfaces composed of 2 monolayers of a 50%-50% PdPt Alloy. These results support three prior examples of agreement between calculations and measurements for pairs of metals having the same crystal structure and the same lattice parameter to within 1%. We also estimate the spin-flipping probability at Pd/Pt interfaces as $\delta_{Pd/Pt}$ = 0.13 ± 0.08.

Recently it has become possible to calculate, with no-free-parameters, the specific resistance of a pair of interfaces, $2AR_{M1/M2}$, of metals M1 and M2 that have the same lattice structure and the same lattice parameter [1-3]. Here R is the current-perpendicular-to-plane (CPP) resistance of the interface and A is the cross-sectional area through which the current flows. In three prior cases, fcc Ag(111)/Au(111) [4], fcc Co(111)/Cu(111) [5], and bcc Fe(110)/Cr(110) [6], calculations assuming flat, perfect interfaces agreed surprisingly well (within 10-20% [3,8]) with experimental values for sputtered multilayers. These values ranged from a low of $2AR_{Ag/Au}$ ~ 0.1 f$\Omega$m$^2$ to a high of $2AR_{Fe/Cr}$ ~ 1.6 f$\Omega$m$^2$. Given the disordered nature of sputtered interfaces, these agreements were surprising. The surprise lessened when calculations showed that the 2ARs for these metal pairs changed only modestly when the flat, perfect interfaces were replaced by disordered ones composed of two interfacial monolayers (ML) of a 50%-50% alloy [3].

In contrast to these agreements, calculations and measurements for metal pairs with the same lattice structures but lattice parameters differing by 5% to 13% disagreed more substantially, in the best case by a factor of 1.5 (Au/Cu [4,8]), but more typically by ~ 2 [7,8], and in the worst case by 4.5 [4,8]. To test the importance of the magnitude of the difference in lattice parameter, results were compared for Pd combined with Cu (lattice parameter smaller than Pd's by 8% [7]) or with Ag or Au (lattice parameter larger than Pd's by only 5% [8]) The 2ARs for Pd/Ag and Pd/Au differed from the calculations by more than that for Pd/Cu. When the lattice parameters of M1 and M2 are not the same, one must choose how to treat the difference in lattice parameters, since self-consistent calculations of relaxation in the vicinity of the interface are not yet feasible. For Pd/Ag, none of a variety of different choices for this difference brought the calculation close to the measured value [8].

Given the less-satisfactory results with pairs of metals having different lattice parameters, we decided that it is important to compare theory and experiment for 2AR with additional pairs having lattice parameters differing by ≤ 1%. Pd and Pt constitute one such pair [9], with the advantages that Pd and Pt both sputter well [7,8], are completely miscible, and do not form intermetallics [10]. Their combination should, thus, minimize unwanted experimental problems. In this letter we report determinations of $2AR_{Pd/Pt}$ using two different techniques, and compare the results with calculations for both perfect and disordered interfaces. One of the techniques lets us also determine an additional interesting parameter, the spin-flipping probability, $\delta_{Pd/Pt}$, at a sputtered Pd/Pt interface,

To find AR for a multilayer, we must measure both A and R. Our sample geometry and techniques of sputtering and measuring resistance are described in [11]. We achieve uniform current flow through A ~ 1.2 mm$^2$ by using crossed superconducting niobium (Nb) strips, 150 nm thick and ~ 1.1 mm wide [11]. Using superconducting Nb limits us to measurements at low temperature (4.2K). A is the product of the widths of the two Nb strips, measured with a Dektak profilometer [11]. Comparisons of multiple measurements, of A, including by different students, yield typical uncertainties of ± 4-5%, with occasional larger ones as shown by error bars in the graphs below. We estimate the resistivities of our Pd and Pt from separate Van der Pauw measurements of sputtered 200 nm thick thin films. For Pd, two samples gave $\rho_{Pd}$ = 46 ± 1 n$\Omega$m, similar to $\rho_{Pd}$ = 40 ± 3 n$\Omega$n from [7]). However for Pt, three samples gave $\rho_{Pt}$ = 26 ± 1 n$\Omega$m, lower than $\rho_{Pt}$ = 42 ± 6 n$\Omega$m from [7]). As the 4.2K resistivity of a sputtered nominally-pure metal is dominated by residual defects, it may well vary between runs made years apart. We use the present values in our data analysis, but include the previous one to estimate uncertainties.

To test for systematic errors in our estimates of $2AR_{Pd/Pt}$, we determined it in two different ways.

The first technique for finding $2AR_{Pd/Pt}$ [4] uses the multilayer Nb(100)/Co(10)/[Pd(t)/Pt(t)]$_N$/Co(10)/Nb(100), where thicknesses are in nm, t is the common thickness of the Pd and Pt layers, $N$ is the number of Pd/Pt bilayers, and the total thickness of the [Pd(t)/Pt(t)]$_N$ multilayer is held fixed at 360 nm. The Co layers kill the proximity effect from the superconducting Nb that would otherwise turn part of the

Pd/Pt multilayer superconducting. The Co layers are far enough apart that sample magnetoresistance is negligible, as checked by measurements to ± 1 kOe. If the Pd and Pt layer resistivities are independent of the layer thickness t, then the two-current series-resistor (2CSR) model [12] predicts a total specific resistance, $AR_T$ of [4]:

$$AR_T = 2AR_{Nb/Co} + 2\rho_{Co}\text{x}10\text{nm} + AR_{Co/Pd} + AR_{Co/Pt} + \rho_{Pd}\text{x}180\text{nm} + \rho_{Pt}\text{x}180\text{nm} + N(2AR_{Pd/Pt}), \quad (1)$$

until t becomes less than the finite thickness $t_I$ of the Pd/Pt interfaces (typically ≤ 1 nm [4]). Eq. (1) can be written as $AR_T = C + 2N(AR_{Pd/Pt})$, where C is the sum of the first 6 terms. We have independently measured all of the parameters in C [5] except for $AR_{Co/Pd}+ AR_{Co/Pt}$. If we use a rough typical value of 1 f$\Omega$m$^2$ for the sum $AR_{Co/Pd}+ AR_{Co/Pt}$, we predict C = 21 ± 4 f$\Omega$m$^2$. We will see below that this range overlaps our experimental fits of 18 ± 1 f$\Omega$m$^2$ for two sets of data. For t << $t_I$, AR should saturate at the product of 360 nm times the resistivity of a 50%-50% Pd-Pt alloy.

The second technique for finding $2AR_{Pd/Pt}$ involves inserting an X = [Pd(3)/Pt(3)]$_N$ multilayer into the middle of an exchange-biased spin-valve of the form Nb(100)/FeMn(8)/Py(24)/Cu(10)/X/Cu(10)/Py(24)/Nb(100) and measuring the increase in AR as N is increased [13]. In these samples, the FeMn exchange-bias pins the adjacent Py layer to a much higher coercive field than the other 'free' Py layer [13,14]. The Py layers are made much larger than the Py spin-diffusion length at 4.2K, $l_{sf}^{Py} \cong 6$ nm [14], both to maximize the CPP-MR and to make the spin-valve properties only weakly dependent on the Py layer thickness. For such a sample, the Valet-Fert [12] extension of the 2CSR model predicts that the average $AR_A = [AR(AP) + AR(P)]/2$ should increase linearly with N proportional to $AR_{Pd/Pt}$ as [8]:

$$AR_A = C' + N[(\rho_{Pt} + \rho_{Pd}) \text{ x } 3 \text{ nm} + 2AR_{Pd/Pt}]. \quad (2)$$

Here C' is the sum of several terms that are independent of N, plus 2 interface terms, $AR_{Pd/Cu}$ and $AR_{Pt/Cu}$, that are present only for N > 0. To eliminate these latter two terms, we omit the data for N = 0 from our fit. Prior measurements give values of C' ranging from 18 to 24 f$\Omega$m$^2$ [7,8,13]

The Valet-Fert model also predicts that the difference in specific resistance, $A\Delta R = AR(AP) – AR(P)$, should decrease exponentially with increasing N as [13]:

$$A\Delta R \propto \exp(-N[2\delta_{Pd/Pt} + t/l_{sf}^{Pd} + t/l_{sf}^{Pt}]). \quad (3)$$

Here, $l_{sf}^{Pd}$ and $l_{sf}^{Pt}$ are the spin-diffusion lengths in Pd and Pt. Because increasing N also increases the amount of Pd and Pt in X, we must use Eqs. (2) and (3) to correct the measured values of both $AR_A$ and $A\Delta R$ for contributions from the 'bulk' of the Pd and Pt. Because $2AR_{Pd/Pt}$ turns out to be relatively small, while $\rho_{Pd}$ and $\rho_{Pt}$ are moderately large, the correction for Pd and Pt is a major fraction of the total, making the value of $2AR_{Pd/Pt}$ derived in this way more uncertain than we would have liked.

Fig. 1 shows $AR_T$ vs N for two sets of multilayers of the first type. The first sample set (circles) was made a while ago by one of us and has been corrected for a programming error that led to an excess layer thickness of ~ 0.5 nm/bilayer. To independently check this correction, another of us recently made a second sample set (squares), using correct programming as established by x-rays. The similarities and differences between the two data sets illustrate our reproducibility over long times. Both sets are consistent with straight lines up to about N = 120, after which they fall below that line. Their common intercept is 18 ± 1 f$\Omega$m$^2$ (both circles and squares), consistent with our estimate above. The slopes of the best linear fits up to N = 120 are 0.27 ± 0.02 f$\Omega$m$^2$ (circles) and 0.32 ± 0.01 f$\Omega$m$^2$ (squares). Averaging these values gives our 'best estimate' for this technique of 0.29 ± 0.03 f$\Omega$m$^2$. As the interfaces begin to overlap, the data should fall below the line associated with Eq. 1 and eventually approach a constant 'saturation value' limit, $AR_S$, associated with a uniform 50%-50% alloy. The data for Pd/Pt alloys at 4.2K in [15] gives a resistivity for a 50%-50% alloy of ~ 21 $\mu\Omega$cm. Multiplying this resistivity by a total sample thickness of 360 nm gives AR(Pd/Pt) ~ 76 f$\Omega$m$^2$. Adding the additional terms of Eq. 1 for N = 0 (neglecting the Pd and Pt terms that are replaced by the alloy) gives an estimate of $AR_S$ ~ 84 f$\Omega$m$^2$, comparable to, but lower than, our actual values at N = 450.

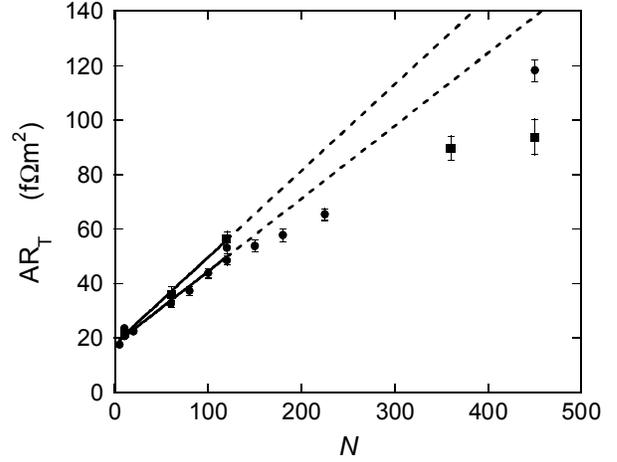

Fig. 1. $AR_T$ vs N for [Pd/Pt]$_N$ multilayers. The solid lines are least-squares fits to the circles and squares up to N = 120, giving $2AR_{Pd/Pt} = 0.32 \pm 0.01$ f$\Omega$m$^2$ (squares) and $0.27 \pm 0.02$ f$\Omega$m$^2$ (circles).

Fig. 2 shows $AR_A$ vs N for the spin-valve-based multilayers. The linear fit shown neglects the data points at N = 0 for the reason given above. The slope of this line is 0.39 ± 0.13 f$\Omega$m$^2$. Subtracting our best value of [$(\rho_{Pt} + \rho_{Pd})$ x 3 nm] = 0.22 ± 0.03 f$\Omega$m$^2$ gives $2AR_{Pd/Pt} = 0.17 \pm 0.13$ f$\Omega$m$^2$. Within its larger uncertainty, this value agrees with that determined by the other technique.

The weighted average of our two values of $2AR_{Pd/Pt}$, gives our best estimate of $2AR_{Pd/Pt} = 0.28 \pm 0.06$ f$\Omega$m$^2$, smaller than the values for Pt/Cu ($2AR_{Pt/Cu}$ ~ 1.5 f$\Omega$m$^2$ [7]), Pd/Cu ($2AR_{Pd/Cu}$ ~ 0.9 f$\Omega$m$^2$ [7], Pd/Ag ($2AR_{Pd/Ag}$ ~ 0.7 f$\Omega$m$^2$[8]), or Pd/Au ($2AR_{Pd/Au}$ ~ 0.45 f$\Omega$m$^2$[8]). However, this value agrees reasonably well with our no-free-parameter calculations of $2AR_{Pd/Pt}$. Applying the local density approximation (LDA), we find for a flat, perfect interface, $2AR_{Pd/Pt}(perf) = 0.30 \pm 0.01$ f$\Omega$m$^2$, and for 2 ML of a 50%-50% alloy, $2AR_{Pd/Pt}(50-50) = 0.33 \pm 0.01$ f$\Omega$m$^2$. The calculated Fermi surfaces for Pd and Pt can derivate from experiment by 5 mRy [16]. Allowing for uncertainties in the Fermi energies gives $2AR_{Pd/Pt}(perf) = 0.30 \pm 0.04$ f$\Omega$m$^2$, and $2AR_{Pd/Pt}(50-50) = 0.33 \pm 0.04$ f$\Omega$m$^2$.

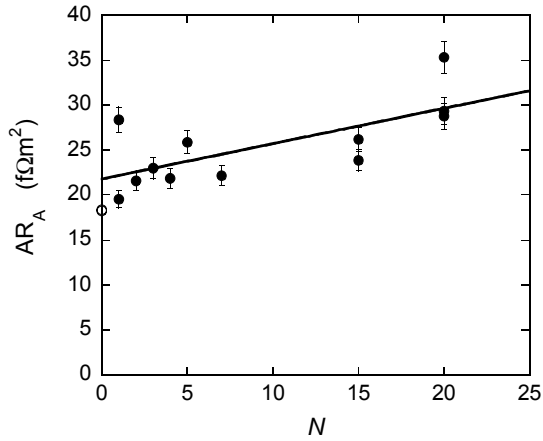

Fig. 2. $AR_A$ vs $N$ for $[Py(3)/Pt(3)]_N$ multilayers embedded inside Py-based EBSVs. The line is a least-squares fit, neglecting the data at $N = 0$.

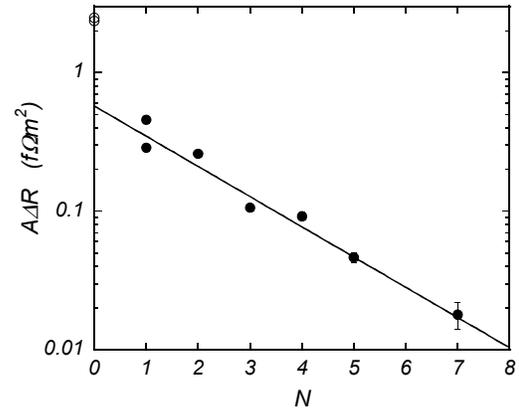

Fig. 3. log $A\Delta R$ vs $N$ for $[Py(3)/Pt(3)]_N$ multilayers embedded inside Py-based EBSVs. The line is a least-squares fit, neglecting the data at $N = 0$.

Turning to spin-flipping at a Pd/Pt interface, Fig. 3 shows $A\Delta R$ vs $N$ for the same samples as in Fig. 2. For the same reason as in Fig. 2, we fit the data neglecting the two points at $N = 0$. The resulting slope is -0.50 ± 0.02. To obtain $\delta_{Pd/Pt}$, we must correct for spin-flipping within the Pd and Pt layers (see [13]). For this correction we take t = 3 nm and use our published value for Pd, $l_{sf}^{Pd} = 25^{+10}_{-4}$ m [7]. Since we expect [12] $l_{sf}^{Pt} \propto \Lambda_{Pt}$, the mean-free-path in Pt, we renormalize our published value for Pt by the ratio of the residual resistivities (see above) in that study and this one to get $l_{sf}^{Pt} = (4.2/2.6)[14 \pm 6$ nm$] = 23 \pm 10$ nm. Combining all these values gives a best estimate of $\delta_{Pd/Pt} = 0.13 \pm 0.08$. This value is below the $\delta \sim 1$ found for the heavy metals W or Pt combined with a light metal like Cu [3], but comparable to the values $\delta \sim 0.1$-0.24 for Pd with Au, Ag, or Cu [7,8].

In summary, we have measured $2AR_{Pd/Pt}$ for sputtered samples in two different ways, deriving a 'best value' of $2AR_{Pd/Pt}(Exp) = 0.28 \pm 0.06$ f$\Omega$m$^2$. This value agrees within mutual uncertainties with no-free-parameters calculations of $2AR_{Pd/Pt}(Calc) = 0.30 \pm 0.04$ f$\Omega$m$^2$ for flat, perfect interfaces and $2AR_{Pd/Pt}(Calc) = 0.33 \pm 0.04$ f$\Omega$m$^2$ for interfaces with 2ML of a 50%-50% alloy. We derive also a spin-flip probability at a sputtered Pd/Pt interface of $\delta_{Pd/Pt} = 0.13 \pm 0.08$.

We acknowledge support from the MSU CFMR, CSM, NSF grants DMR 02-02476 and 98-09688, Seagate Technology, and NSF of China Grant 90303014/A0402.